\documentstyle[11pt,aaspp4]{article}

\def\h2co{\hbox{H$_2$CO}}

\def\H76{\hbox{H~76$\alpha$}}
\def\H2O{\hbox{H$_2$O}}
\def\deg{\ifmmode^\circ\else$^\circ$\fi}
\def\degns{\ifmmode^\circ\else$^\circ$\fi}

\def\arcs{\char'175~}

\def\kms{~km~s$^{-1}$}

\def\eg{e.g.,~}

\def\etal{et~al.~}

\def\NH3{\hbox{NH$_3$}}
\def\vlsr{\hbox{V$_{lsr}$}}
\def\kms{km s$^{-1}$}

\def\ts{\thinspace}
\def\w44{W\thinspace{44}}
\def\ic443{IC\thinspace{443}}

\begin{document}

\title {Polarization Observations of 1720 MHz OH Masers toward 
the Three Supernova Remnants W~28, W~44, and IC~443}

\author{M. J Claussen, D. A. Frail, and W. M. Goss}
\affil{National Radio Astronomy Observatory, P.O. Box 0, 
Socorro, New Mexico 87801}
\centerline {and} 
\author{R. A. Gaume}
\affil{United States Naval Observatory, 3450 Mass. Ave., Washington,
DC 20392}

\vfill\eject

\begin{abstract}

We present arcsecond resolution observations from the VLA\footnote{The
Very Large Array (VLA) is operated by NRAO, a facility of the National Science
Foundation operated under cooperative agreement by Associated
Universities, Inc.} with full Stokes polarimetry of the ground-state
satellite line of the hydroxyl molecule (OH) at 1720.53 MHz ($^2\Pi_{3
\over 2}, J = {3 \over 2}, F = 2 \rightarrow 1$) toward three Galactic
supernova remnants: W\thinspace{28}, \w44, and \ic443. The total
number of individual OH(1720 MHz) ``spots'' we detect in each of these
three remnants is 41, 25 and 6, respectively. The OH(1720 MHz)
features appear to lie along the edge of radio continuum emission from
the supernova remnants, but are displaced behind the leading edge of
the shock as traced by the synchrotron emission. The brightness
temperatures of the OH(1720 MHz) emission features range from
2$\times$10$^4$ to 10$^8$ K, convincingly demonstrating the maser
nature of the OH(1720 MHz) features. We argue that the partially resolved
angular diameters that we measure for the masers are neither intrinsic
sizes nor scattering disks, but result from a blend of several
unresolved maser features near the same velocity. Thus our computed
brightness temperatures are lower limits to the true values. The
characteristic antisymmetric ``S'' profile, indicative of Zeeman
splitting in the weak-field case, is identified in the Stokes {\bf V}
spectrum of several of the brighter maser spots. The derived
line-of-sight magnetic fields are of order 0.2 mG and are remarkably
constant in both direction and in magnitude over regions several
parsecs apart. These are the first measurements of post-shock magnetic
fields in supernova remnants and demonstrate the importance of
magnetic pressure in these molecular shocks. The velocity dispersion
of the maser features is typically less than a few km
s$^{-1}$, and except in the special case of W\thinspace{28}, the mean
maser velocity is equal to the systemic velocity of the remnant.
We suggest that the maximum amplification of the maser transition will 
occur when the acceleration produced by the shock is transverse to the 
line of sight. Additional support for this point comes from the location 
of the masers in \ic443, and molecular observations which allow the shock
geometry to be determined. All of our observations are consistent with 
a model in which the OH(1720 MHz) is collisionally excited by H$_2$ 
molecules in the postshock gas heated by a non-dissociative shock. 
Finally, we end with a discussion of the importance of supernova 
remnants with OH(1720 MHz) maser emission as promising candidates to 
conduct high energy searches for the sites of cosmic ray acceleration.

\end{abstract}


\vfill\eject

\section{Introduction}

Recognition of the importance of the ground-state satellite line of
the hydroxyl molecule (OH) at 1720.53 MHz ($^2\Pi_{3 \over 2}, J = {3
\over 2}, F = 2 \rightarrow 1$) as a powerful tool for studying the
interaction of supernova remnants (SNRs) with molecular clouds has
been slow in coming. Anomalous OH(1720 MHz) emission toward the SNRs
W\ts{28} and \w44 was originally noted by Goss (1968), Ball \& Staelin
(1968) and Turner (1969).  These and follow up observations by Goss \&
Robinson (1968) showed that the OH(1720~MHz) line in these regions was
strong and narrow and did not have accompanying emission at the other
ground-state transitions of OH (1612, 1665 and 1667 MHz), which were
instead seen in absorption. Following this initial discovery, other
authors noted the OH(1720 MHz) line toward W\ts{28}, \w44, and \ic443\
(\eg Hardebeck 1971, Haynes \& Caswell 1977, DeNoyer 1979, Turner
1982), but little or no investigation was made as to whether the
OH(1720 MHz) lines were associated with the supernova remnants, a
suggestion that was tentatively made by Goss \& Robinson (1968). The
first evidence that this may be the case was presented by Frail, Goss,
\& Slysh (1994a), who observed W\ts{28} at 12\arcs resolution 
with the VLA. A
total of 26 OH(1720 MHz) emission ``spots'' were found distributed
along the edge of the remnant. The compact nature and high brightness
temperature (T$_b\sim{4}\times{10}^{5}$ K) of the spots suggested that
these were masers. Based on the location of the masers, and the
physical properties of the supernova remnant and the adjacent
molecular cloud, Frail et al. (1994a) argued that the OH(1720 MHz)
masers in W\ts{28} were consistent with a shock excitation model
advocated by Elitzur (1976).

This effort has continued in earnest with a large-scale SNR survey for
OH(1720 MHz) with the Green Bank 43-m and Parkes 64-m telescopes, and
confirmation observations carried out at the VLA and the Australia
Telescope Compact Array (Frail et al. 1996, Green et al. 1997). Based
on the early success of identifying OH(1720 MHz) emission toward SNRs
near the Galactic Center (Yusef-Zadeh, Uchida \& Roberts 1995,
Yusef-Zadeh et al. 1996) a VLA survey was conducted of a much larger
sample (Robinson et al. 1997). In total we have looked for this line
from $\sim$166 SNRs and OH(1720 MHz) lines have been confirmed from 17
SNRs, for a detection rate of 10\%.

Somewhat less effort has been made on the theoretical side. Most of
the recent work concerning the excitation of the OH molecule has been
for masers in star-forming regions (Cesaroni \& Walmsley 1991, Gray,
Doel, \& Field 1991). Here the strong far infrared (FIR) radiation
field from the dust in the HII region virtually assures that the
dominant pumping mechanism is radiative not collisional.  The only
work that specifically explores the excitation of OH(1720 MHz) under
the conditions we expect for SNRs is Elitzur (1976). He showed that in
order to create a strong ($-\tau>>1$) inversion of the 1720 MHz line,
collisions of the OH molecule with H$_2$ had to occur in a range of
kinetic temperature and density between 25 K$\leq$T$_k\leq$200 K and
10$^3$\ cm$^{-3}\leq{n_{\rm{H}_2}} \leq{10}^5$\ cm$^{-3}$,
respectively. Pavlakis \& Kylafis (1996a, 1996b) have recently
re-visited this topic using newer collisional rates between OH and
H$_2$ (Offer, van Hemert \& van Dishoeck 1994) and they included a
variety of different pumping mechanisms (collisions, FIR field, local
and non-local line overlap). Considering only collisions and local
line overlap mechanisms over a restricted range of kinetic temperatures
(T$_k$=100-200 K), they find that bright OH(1720 MHz) maser emission
T$_b\simeq 10^9$ K results at densities (within a factor of three) of
${n_{\rm{H}_2}}=3\times{10}^{5}(10^{-5}/f_{OH})$ cm$^{-3}$, where
$f_{OH}$ is the fractional abundance of OH relative to H$_2$. Pavlakis
\& Kylafis (1996a) also note that OH(1720 MHz) is a sensitive
indicator of the amount of ortho-H$_2$ present relative to para-H$_2$,
with the inversion reaching a maximum when ortho-H$_2$ dominates. With
further modeling of this type it should be possible to infer the
physical conditions of the post-shock gas from the OH(1720 MHz) masers
alone.

All the observational work to date has been concerned primarily with
identifying new examples of SNR's with the OH(1720 MHz) line. The full
potential of the OH(1720 MHz) has yet to be utilized. In this paper we
present a detailed study of three SNRs: W\ts{28}, \w44, and
\ic443. Together they constitute the three best examples of SNRs
interacting with molecular clouds. Observations of molecular
transitions in the millimeter-wavelength regime toward these three
SNRs (\eg Wootten 1977, Pastchenko \& Slysh 1974,
DeNoyer 1979, Wootten 1981, Dickman \etal 1992) provide convincing
evidence that an interaction is taking place. The goal of the
observations reported in this paper are (1) to demonstrate that the OH
emission regions are indeed masers by measuring the brightness
temperature of the OH(1720 MHz) spots at arcsecond resolution, (2) to
demonstrate that the polarized emission from the OH(1720 MHz) masers
is due to the Zeeman effect and to measure the magnetic field at
several locations in the post-shock gas, (3) to infer physical
properties of the environment of the molecular cloud/SNR interface
based on the models for the maser emission mechanism and other
molecular lines, (4) determine the angular sizes of the masers and to
establish whether they are intrinsic or due to strong scattering, and
(5) to discern the velocity structure of the interface by observing
the spatial distribution of different velocity components.

\section{Observations}

Observations of the 1720 MHz OH line in three supernova remnants,
W\ts{28}, \w44\ and \ic443\ were conducted on March 12 and March 17,
1994, using the Very Large Array of the National Radio Astronomy
Observatory.\footnote{The National Radio Astronomy Observatory is a
facility of the National Science Foundation, operated under a
cooperative agreement by Associated Universities, Inc.}  The VLA was
in the {\bf A} configuration, so the resolution at $\lambda$=18 cm was
about 1\arcs.  Variations in the {\it u,v} coverage for observations
of the three supernova remnants make the resulting synthesized beam
somewhat different.  The synthesized beam sizes for each source are
given in Table 1 with the major axis (in arcseconds), the minor axis
(in arcseconds) and the rotation of the beam from North in a
counterclockwise direction (in degrees).  The VLA correlator was
configured to produce all four parallel- and cross-polarization
products, which were processed into the Stokes parameters {\bf I, Q,
U} and {\bf V}. The data were Hanning smoothed on-line to give
velocity resolution of 0.53 \kms.  The total bandwidth was 195 kHz,
providing a total velocity coverage of 33.5 \kms.  For W\ts{28}, three
pointing centers and different central velocities were used in order
to observe all sources of the 1720 MHz masers as discovered by Frail
\etal (1994a).  For \w44\ and \ic443\ only one pointing center for
each supernova remnant was required.  Table 1 also gives the position
of the pointing centers and the central velocities of the
observations.  The data for each pointing center were calibrated in
the standard manner, including polarization calibration.  Images were
produced by natural weighting of the {\it u,v} data, and were CLEANed
in order to remove the effects of the incomplete {\it u,v} coverage.
For some pointings, multiple fields within the primary beam of the
antenna were imaged and cleaned.  Images were made in all four Stokes
parameters.  The rms noise per spectral channel obtained for each
object is summarized in the final column of Table 1.

After the imaging step, individual maser features were identified.
This was done by carefully examining each spectral channel image
individually, and then using the {\bf AIPS} tasks {\bf JMFIT} and {\bf
SAD} to fit the positions of emission features.  The positions
resulting from using these two tasks were compared and found to agree
within the estimated fitting errors (which, for features of typical
strength, are about 0.01 times the synthesized beam).  For weak
features ($<$ 100 mJy beam$^{-1}$) the fitting errors are somewhat
larger, up to 0.1 times the synthesized beam.  Positions of individual
features were measured from each spectral channel where emission was
noted.  Gaussians were fit in the spectral domain when maser features
at the same spatial position appeared in three or more adjacent
spectral channels. The absolute positions of the maser features are
tied to the positions of the phase calibrators 0629+104 (\ic443),
1748$-$253 (W\thinspace{28}), and 1801+010 (\w44). None of these are
astrometric-quality calibrators, so while the {\it relative} positions
of the brightest masers are limited by signal-to-noise to a few 10's
of milliarcseconds, the {\it absolute} positions are probably no
better than $\pm{100}$ mas.

The angular sizes of the maser features were also measured.  For this
data, where multiple non-overlapping fields were mapped from a single
{\it u,v} data set, and several maser features are typically found in
each field, it is very difficult to measure sizes of individual
features by fitting to the {\it u,v} data.  Instead, we fit
two-dimensional Gaussians to individual maser features and the sizes
were deconvolved from the synthesized beam using the AIPS task {\bf
JMFIT}.  Tables 2, 3, and 4 show the results of these fits in the
spatial and spectral domain for W\ts{28}, \w44, and \ic443,
respectively.

\section{Results}

\subsection{Distribution of Masers}

In Figures 1, 2, and 3 we show images of the radio continuum for the
SNR W\ts{28}, \w44, and \ic443, respectively. Rather than show the
positions of individual features from Tables 2, 3 and 4 we show {\it
regions} of OH(1720 MHz) concentrations overlaid on the grey-scales of
the SNRs. Since the total size of the remnants is more than 30
arcminutes and we observed maser emission with $\sim$1.5\arcs\
resolution, we must show these regions separately in order to
distinguish individual maser features. These are given in Figures 4-6.

For W\ts{28} the apparent concentration of masers along the eastern
and north-central part of the remnant is real. With our three
pointings we should have been capable of detecting bright masers all
across the face of W\ts{28}. The detections are presumably telling us
about the extent of the interaction of the shock with the adjacent
molecular cloud. Comparing these findings with Frail et al. (1994a) we
detect 41 individual features versus their 26 detections.  All of the
bright masers detected by Frail et al. (1994a) are seen in our data but
we fail to detect some of their weaker masers ($<100$ mJy) presumably
due to our higher noise level. In Figure 4a-4f we show the expanded
regions of W\ts{28}, with the position of unresolved maser features
marked by filled circles and resolved features by ellipses. The order of
magnitude improvement in resolution between these and the earlier
W\ts{28} data shows that several of the original spots were in fact
blended features. This is most dramatic in the region denoted as
W\ts{28}-OH\ts{E} (see Fig. 4e). Instead of the four masers detected
at 15\arcs\ resolution by Frail et al. (1994a) there are at least two
dozen separate masers in this region at 1.5\arcs\ resolution.

Similar numbers of masers (25) are found for \w44\ but in \ic443\ we
discovered only six masers. A possible explanation for the relative
paucity of masers in \ic443\ will be discussed in \S{4.3}. Like
W\ts{28}, the masers in \w44\ are concentrated in several regions,
with the largest number of masers located near where the densest part
of the molecular cloud is in physical contact with the remnant
(Wootten 1977). The masers in all three remnants appear along the
continuum edges of the synchrotron emission (see Figs. 4-6). This is
most apparent in W\ts{28}-OH\ts{E} and W\ts{44}-OH\ts{E}, where the
masers are plentiful enough to show that they lie along a line traced
by the continuum contours. OH main-line masers have been found in
linear arrangements before. W75N is a notable example with a N-S line
of OH masers and a velocity spread of 9 km s$^{-1}$ over its full
1.5\arcs\ extent (Haschick et al. 1981, Baart et al. 1986). These
observations have been interpreted alternatively as masers
delineating a rotating protostellar disk (Haschick et al. 1981), or
masers along the edge of an HII region, in gas compressed by an
expanding shock (Baart et al. 1986). For the OH(1720 MHz) masers being
discussed here, their close location near the edge of the non-thermal
radio continuum emission is suggestive. The synchrotron emission
results from particle acceleration in a shock, either in the outer
blast wave or the reverse shock propagating into the ejecta (Blandford
\& Eichler 1987). Thus we favor an interpretation where the OH(1720 MHz)
masers are located in the shock-compressed gas behind this region.

\subsection{Maser Kinematics}

The velocity of the masers in \w44\ and \ic443\ have a low dispersion
($<$1-2 km s$^{-1}$) about a mean value that agrees with the systemic
velocity of the remnant (see \S{4.2} and {4.3}). This point was noted
earlier for the SNRs studied by Frail et al. (1996) and in fact they
concluded that the velocity of the OH(1720 MHz) masers could be used
to derive a kinematic distance to the remnants. The masers in W\ts{28}
appear to be an exception to this rule. There is a broad distribution of maser
velocities in Table 2, ranging from 4.8 km s$^{-1}$ in
W\ts{28}-OH\ts{A} to 16 km s$^{-1}$ in W\ts{28}-OH\ts{E}. The velocity
dispersion of the masers within individual regions is small, typically
1-2 km s$^{-1}$. On the larger scale there are large velocity
difference across the remnant, reaching a maximum at W\ts{28}-OH\ts{E}
and falling away in regions both to the north and south.

We plot the velocity of these maser features as a function of angular
distance from the remnant center in Figure 7.  Velocities of maser
features which lie closest to the center of W\thinspace{28} are the
smallest. There is a gradual increase from \vlsr=5 km s$^{-1}$ at
$\theta=3^\prime$ to \vlsr=8 km s$^{-1}$ at $\theta=14.5^\prime$.  At
the edge of the remnant ($\theta=14.5^\prime$) the velocity dispersion
increases sharply.  This confluence of large positive velocities near
the region where W\ts{28} is in direct contact with the molecular
cloud as mapped by Wootten (1981) was noted by Frail et
al. (1994a). However, it is clear from Figure 7 that a simple 
geometric model 
such as a tilted disk or expanding sphere does not fit the velocity
profile. We will return to this point in \S{4.1}.

\subsection{Angular Diameters of the Maser Spots}

The brightness temperatures calculated for the OH(1720 MHz) emission
features in Tables 2-4 range from about 2 $\times10^4$ to 10$^8$ K.
This result, together with their narrow linewidths, leaves little
doubt that the emission is amplified, stimulated emission
(i.e. masers). Even the weakest emission sources have brightness
temperatures much larger than expected in the molecular environment.
We estimate that the continuum emission from the three SNRs, which are
acting as background sources for the maser amplification process, is
10-80~K. Based on the brightness temperatures given above, this
implies lower limits to the optical depths $\tau$ for the OH(1720 MHz)
lines of $-8$ to $-16$.

There is a range of apparent angular sizes for the maser features in
Tables 2-4. Our deconvolution process has measured spot sizes on the
order of 0.5\arcsec-0.9\arcsec\ for a number of features, and still
other features are distinctly larger than the synthesized
beam. Several maser spots are non-circular with aspect ratios larger
than that of the synthesized beam. The
orientation of the major axis of these non-circular features
(Figs. 4-6) show no apparent pattern; they are neither aligned with
each other nor do they trace the continuum contours. At the
distance of these remnants the ``resolved'' maser spots
($\theta>2$\arcsec) correspond to linear sizes of $\sim{10}^{17}$
cm. Three possible origins exist for the spot sizes that we see: 1) they
may be intrinsic maser spot sizes, 2) they may be broadened by multipath
scattering in the ionized medium along the line of sight, or 3) they are
due to multiple maser features within the synthesized beam. We favor
the third explanation and we argue this point below.

The apparent maser spot sizes in Tables 2-4 are well in excess of the
expected intrinsic sizes. Observations of the OH(1720 MHz) masers in
the compact HII region W3(OH) give sizes less than 1.2 mas (Masheder
et al. 1994). The theory for the inversion of the OH(1720~MHz) line
predicts $\tau\sim-20$ for the conditions we expect in these remnants
(Elitzur 1976). Converting this to a angular size using the background
continuum levels discussed above, we estimate intrinsic angular sizes
between 1 and 100 mas. Both of these estimates of intrinsic size are
well below the angular diameters that we observe. Some angular
broadening of the intrinsic maser sizes may be expected from
interstellar scattering. In fact, Gwinn, Bartel \& Cordes (1993)
report an excess of scattering for the Vela and Crab pulsars,
attributed to their respective SNRs and Spangler, Fey \& Cordes (1987,
and references therein) note several examples of enhanced scattering
of sources along lines of sight that pass near SNRs. However, the
amount of scattering that the angular diameters in Tables 2-4 imply is
excessive, exceeding the magnitude seen towards such well-known lines
of sight as the Galactic center (Frail et al. 1994b). Furthermore,
there is no indication of excess scattering in the pulse profiles of
PSR\thinspace{B0611+22} and PSR\thinspace{B1853+01} (Taylor,
Manchester \& Lyne 1993), the pulsars which are either behind or in
\ic443\ and \w44, respectively. The pulsar PSR\thinspace{1758}$-23$
toward W\thinspace{28} is a special case and does show a broad
scattering tail in its pulse profile.  From the amount of temporal
broadening in PSR\thinspace{1758}$-23$ and upper limits to the angular
diameter of a nearby extragalactic source, Frail, Kulkarni \& Vasisht
(1993) constrain the SNR-scattering screen geometry and give a
prediction for the angular diameter of point sources seen at
W\thinspace{28}. At 1720 MHz the angular broadening in W\thinspace{28}
is expected to be between 150 and 500 mas. This is large but again well below
the values we measure here. The simplest explanation consistent with
the available data is that the elongated features that we see are a
blend of unresolved maser features near the same velocity. Thus our
computed brightness temperatures are {\it lower limits} to the true
brightness temperatures. We have already seen how increasing the
resolution by an order of magnitude in W\thinspace{28} (15\arcsec\ to
1.5\arcsec) resulted in the discovery of many more features. If we are
correct, then a similar increase should be seen when higher resolution
observations are made.


\subsection{Zeeman Splitting and Magnetic Fields}

Since we recorded all four parallel- and cross-polarization products
from each antenna pair, we can calculate the full set of Stokes
parameters ({\bf I, Q, U, V}) for each line.  Approximately half of
our maser features show significant signal in the Stokes {\bf V}
profiles.  The number of features detected in the {\bf V} images is
probably limited due to sensitivity effects, since the features with
the weakest {\bf I} signal are the ones in which we do not detect
circularly polarized emission. When the polarized signal is located in
the {\bf V} images it is found to coincide on the sky to within
one-tenth of a synthesized beam ($\sim{5}\times{10}^{15}$ cm) to the
emission on the corresponding {\bf I} images. This positional
coincidence is a strong argument that the {\bf V} signal is due to the
Zeeman effect.  Many of the {\bf V} profiles detected have the
classical antisymmetric ``S'' shape about the line center. This also
strongly supports the Zeeman splitting interpretation.

When the Zeeman splitting is small compared to the intrinsic (Doppler)
linewidth, the Stokes {\bf V} profile is proportional to the frequency
derivative of the Stokes {\bf I} profile (Heiles et al. 1993). Thus a
fit of the {\bf V} profile to the derivative of the {\bf I} profile
yields a measurement of the line-of-sight magnetic field,
V=C\thinspace{dI/d$\nu$}, where C=0.6536\thinspace{B$_\parallel$} Hz
$\mu$G$^{-1}$. The proportionality constant that we use is the same as
Yusef-Zadeh et al. (1996) since our definitions of {\bf I} and {\bf V}
are identical. Figures 8 and 9 show sample results of the Stokes
and derivative spectra for maser features in W\thinspace{28}
and \w44, respectively.  In \ic443\ we detect none of the typical
``S'' profiles in {\bf V} and thus a line-of-sight field cannot be
estimated in the same manner. Table 5 lists the results of the Stokes
{\bf V} fitting for the features in all three supernova remnants and
when possible, the values of B$_\parallel$. In both W\thinspace{28}
and \w44\ the strength of the field is within a factor of three of
B$_\parallel$=0.2 mG at {\it all} locations. The direction of the
field throughout each remnant is also constant, implying the existence
of well-ordered magnetic fields over pc-sized distances.

Strictly speaking, the relation used above is valid only for thermal
absorption and emission lines.  The interpretation of polarization 
measurements from maser observations is dependent on the specific
model adopted for polarized radiative transfer.  Nedoluha \& Watson
(1992) conclude that the standard thermal relationship is an excellent
approximation of B$_\parallel$ for observations of water masers, if they
are not strongly saturated, despite the complications of the maser
radiative transfer.  

Elitzur (1996, 1997) has derived a general polarization
solution for arbitrary Zeeman splitting, which reproduces the limits of
zero and very large splitting (compared to the linewidths) studied
by Goldreich \etal (1973) as long as the masers are in the
fully saturated regime.  In particular, Elitzur's solutions span
the range where the Zeeman splitting is comparable to or slightly
less than the linewidth.  For OH, this condition occurs at field
strengths of a few milliGauss.  The solution in this regime has
a similar functional form to that of thermal emission, but the
dependence of the magnitude of Stokes V on the angle ($\theta$) between the prop
agation
direction and the magnetic field direction is somewhat different.
According to Elitzur (1997), masers require smaller fields to
produce the same amount of circular polarization as thermal
emission.  Thus our estimate of the magnetic fields given in
Table 5 may be overestimates, by as much as factors of 2 to 4, depending
on whether or not the 1720 MHz masers are saturated.
These estimates of B$_\parallel$ are five to ten times smaller
than that found near the Galactic center by Yusef-Zadeh \etal (1996).
 
Table 5 also lists the peaks of Stokes parameters {\bf Q} and {\bf U},
the derived fractional linear polarization
(P$_{l} = {{\sqrt{U^2+Q^2}} \over {I}}$)
and the polarization position angle $\chi = 0.5\times tan^{-1} {U\over Q}$,
where they were detected for all three remnants.
According to Elitzur
(1997), for the case of
small splitting compared to the Doppler width, $\theta$, the angle
between the propagation direction and the magnetic field can be
determined directly from a measurement of the linear polarization in
the absence of Faraday rotation.  
Nedoluha and Watson (1990) suggest that the amount of linear
polarization observed must depend on the saturation regime in which
the masers are to be found, and indeed the linear polarization results of 
Goldreich \etal (1973), are also based on how saturated the masers are.
From our current observations, we have no independent measure of the
saturation of the masers, and 
so we have not used the linear polarization measurements
to derive $\theta$.  The relatively small ($\le$ 10\%) linear
polarization that we measure in these observations probably
suggests that the masers are not operating in the completely
saturated regime (Nedoluha \& Watson 1990).

\section{Discussion}

\subsection{W\thinspace{28} (G\thinspace{6.4}$-$0.1)}

Wootten (1981) has summarized the observational evidence that W\ts{28}
is interacting with a molecular cloud.  The case rests on
morphological and kinematic signatures. DeNoyer (1983) looked for, but
did not find, evidence for a shock in the form of abundance
enhancements of various species. On the eastern edge of W\ts{28} there
is a 5$^\prime\times{15}^\prime$ cloud that can be traced in
absorption in the OH(1667 MHz) main-line (Pastchenko \& Slysh 1974)
and H$_2$CO at 4.8 GHz (Slysh et al. 1980), as well as in emission
from $^{12}$CO(1-0) and $^{13}$CO(1-0) (Wootten 1981). This
crescent-shaped cloud at \vlsr=7 km s$^{-1}$ bends around the SNR
along its eastern edge and appears to make contact with W\ts{28} at a
prominent ``kink'' in the radio continuum emission (Velusamy
1987). The line widths of several molecular species (HCO$^+$, CO,
H$_2$CO) broaden significantly in the 7 km s$^{-1}$ cloud as one
moves toward the SNR.  This is nicely shown in Slysh et al. (1980) for
the H$_2$CO absorption. In addition, the gas density, as traced by
$^{13}$CO(1-0) and HCO$^+$, peaks near this contact point (Wootten
1981). The average density in the cloud, determined from H$_2$CO, is
2.5$\times{10}^4$ cm$^{-3}$ (Wootten 1981), although regions of much
higher density are implied by the detection of HCO$^+$. The pre-shock
kinetic temperature is likely T$_k\sim$15 K (DeNoyer 1983).

Together, the morphology and kinematics of this 7 km s$^{-1}$ cloud
has prompted several of the above authors to argue that its
characteristics were being shaped by the impact of the shock from the
W\ts{28} SNR. However, the systemic velocity of W\ts{28} appears to be
much larger than this value. HI absorption measurements by
Radhakrishnan et al. (1972) detect two discrete features at 7.3 km
s$^{-1}$ and 17.6 km s$^{-1}$ against W\ts{28}. Similar lines are seen
in H$_2$CO and OH (Pastchenko \& Slysh 1974, Slysh et al. 1980). The
velocity centroid determined from H$\alpha$ filaments in
W\ts{28} is 18$\pm$5 km s$^{-1}$ (Lozinskaya 1974). If the systemic
velocity of W\ts{28} is close to 17.6 km s$^{-1}$, as the absorption
and emission data suggests, then what is the relationship of the 7 km
s$^{-1}$ molecular cloud to the SNR and what is the source of the
broad distribution in the maser velocities as discussed in \S{3.2}?
We suggest that the entire 7 km s$^{-1}$ molecular cloud
(M=3.2$\times{10}^3$ M$_\odot$) has been accelerated by the SNR shock
along our line of sight, giving it a velocity of $\sim-$10 km s$^{-1}$
from the systemic velocity of W\thinspace{28}. The masers are
collisionally excited by H$_2$ molecules in this cloud as well as
other clouds with no blueshifted velocity component along the
line-of-sight. This may be the origin of the broad range of maser
velocities that we see in W\thinspace{28}, in contrast with all the
other SNRs we have studied to date.

\subsection{W\thinspace{44} (G\thinspace{34.7}$-$0.4)}

\w44\ lies at the base of the Aquila supershell
(GS\thinspace{034}$-$06+65), and is probably one of many SNRs driving
its expansion (Maciejewski et al. 1996). The wealth of observations at
all wavebands has recently been summarized by Giacani et
al. (1997). Here we will review the evidence that the SNR is
interacting with a molecular cloud. Wootten (1977) was the first to
suggest that this was the case on the basis on $^{12}$CO(1-0) and
$^{13}$CO(1-0)CO observations. Near the eastern edge of the remnant
the CO line widths, column density and temperature all
increase. Moreover, he interpreted a velocity gradient seen in the CO
lines as being due to a massive shell (M=2.5$\times{10}^4$ M$_\odot$)
expanding at 4 km s$^{-1}$ toward the observer. In H$_2$CO absorption
away from the remnant Slysh et al. (1980) observe gas at the systemic
velocity of \w44\ (45 km s$^{-1}$) but see a velocity discontinuity as
they move inward along the eastern continuum edge
($\Delta$V$\simeq{5}$ km s$^{-1}$).  This was also interpreted as a
cloud accelerated by \w44. A high-velocity HI shell centered on \w44\
was found by Koo \& Heiles (1995). While they argue that this shell
has been accelerated by \w44\ it lies interior to the outer edge of
the SNR as defined by the continuum emission. Koo \& Heiles develop a
model in which the HI shell was produced by mass outflow from the
progenitor star {\it prior} to the supernova event, and thus is not
part of the current interaction with the ambient molecular gas.

As in the case of W\thinspace{28}, DeNoyer (1983) could not find any
evidence for shock enhancement of molecular species (e.g. CS, HCN,
HCO$^+$) toward \w44.  She further questions the interpretation of the
broad line widths and velocity gradients seen by Wootten (1977) and
Slysh et al. (1980) and suggests that rather than being evidence for
acceleration, they are due to separate but overlapping velocity
components. Perhaps the strongest independent evidence for an
interaction of \w44\ with a molecular cloud comes from recent ISO
observations of [OI], an important coolant in shocked molecular gas
(Reach \& Rho 1996). It was found that the [OI] was brightest along
the eastern edge of the remnants, and in fact peaked at the location
of the OH(1720 MHz) masers. A J-shock model predicting the intensity
of the [OI] line constrains the product of the density and shock
velocity to be 10$^5$ cm$^{-3}$ km s$^{-1}$.

In \S{3.4} we showed that the line-of-sight magnetic field
B$_\parallel$ in both \w44\ and W\thinspace{28} was of the order of
0.2 mG. Recall that this may be an upper limit in B$_\parallel$
because we have used the thermal constant to relate the splitting that
is observed in the Stokes {\bf V} profiles to the magnitude of the
field. By the same token the expectation value of the {\it total}
magnetic field B=2$\times$B$_\parallel$. Recognizing this uncertainty
in the discussion that follows we adopt B=0.2 mG, a value which is
consistent with typical magnetic fields in molecular clouds of density
$n_{\rm{H}_2}=10^4$ cm$^{-3}$ (Myers \& Goodman 1988). The Alfv\'en
velocity V$_A$ in this gas is 4 km s$^{-1}$.

A magnetic field of 0.2 mG has a corresponding magnetic field pressure
p$_B$=B$^2$/8$\pi$ of 2$\times10^{-9}$ dyne cm$^{-2}$. This can be
compared to the thermal gas pressure
p$_g=2\thinspace{n_e}$\thinspace{kT} driving the expansion, as
estimated from the hot X-ray emitting gas filling the interior of the
remnants. In \w44\ Jones, Smith and Angellini (1993) derive
p$_g=6-8\times10^{-10}$ dyne cm$^{-2}$, while for W\thinspace{28}
p$_g\sim{6}\times10^{-10}$ dyne cm$^{-2}$ (Rho et al. 1996). Both of
these values are much larger than the ambient ISM pressure of
p$_\circ={5}\times10^{-13}$ dyne cm$^{-2}$ (Kulkarni \& Heiles 1988).
We conclude on the basis of these estimates that 
p$_B\ge$p$_g\gg$p$_\circ$ and thus the magnetic field exerts
considerable influence on both the dynamics and the structure of the
shock.

The presence of a magnetic field in the molecular gas limits the
postshock compression to $\sqrt{2}$V$_s$/V$_A$, where V$_s$ is the
shock velocity (Draine \& McKee 1993). Lozinskaya (1974) has measured
a mean V$_s$ for the H$\alpha$ filaments in W\thinspace{28} of 45 km
s$^{-1}$, while optical line ratios (Bohigas et al. 1983, Long et
al. 1991) yield values that are typically twice this. In the younger
remnant \w44\ Koo \& Heiles (1995) derive V$_s$=330 km s$^{-1}$ from
modeling HI data, while Rho et al. (1994) derive V$_s$=630 km s$^{-1}$
from X-ray observations. Since these estimates are based on
measurements made in neutral and ionized gas with densities between 1
and 100 cm$^{-3}$, we must scale them by the square root of the
density ratio to obtain a value for V$_s$ where the OH(1720 MHz) maser
emission is seen. The resulting values (assuming
$n_{\rm{H}_2}\sim10^4$ cm$^{-3}$) are highly uncertain but do not
exceed V$_s$=10 km s$^{-1}$. Thus the maximum compression of the gas
and magnetic field in the post-shock molecular cloud is of order two to
three times the ambient value. A related issue is the question of the nature
of the shock. Given the low value of V$_s$ derived above and the
relative strength of the magnetic pressure compared to the thermal gas
pressure (i.e. p$_B\ge$p$_g$), we think it unlikely that the shock
will dissociate the H$_2$ molecules. Whether pre- and post-shock
conditions change gradually (C-type) or suddenly (J-type) cannot be
answered without knowledge of the degree of ionization in the
molecular cloud ahead of the shock (Draine \& McKee 1993). This issue
has considerable bearing on the type of chemistry that occurs behind
the shock and as a result the relative abundances of various
molecules, including OH.



\subsection{IC\thinspace{443} (G\thinspace{189.1}+3.0)}

Since DeNoyer (1979) first found shocked OH in absorption against
IC\thinspace{443}, it has become the best laboratory we have for the
study of the interaction between a supernova remnant shock and a
molecular cloud. At $(l,b)=(189.1,+3.0)$ the physical conditions of
both the pre-shock and post-shock gas for IC\thinspace{443} can be
determined without the usual confusion from unrelated galactic
emission. The neutral hydrogen (HI) and dust were studied by Braun \&
Strom (1986), revealing swept-out cavities, shock heated dust and
accelerated HI gas. Shocked molecular gas along the S-shaped ridge
demarcating the interaction, was observed in the vibrational lines of
H$_2$ by Burton et al. (1988). Numerous studies have been made of
molecules at millimeter and submillimeter wavelengths (e.g. Huang,
Dickman \& Snell 1986, White et al. 1987, Wang \& Scoville 1992, van
Dishoeck, Jansen \& Phillips 1993). Some evidence exists for shock
chemistry processing (i.e. depletions or enhancements for the
abundances of various molecular species relative to their interstellar
values) but not always in accordance with existing shock models. Part
of the difficulty may be that IC\thinspace{443} cannot be modeled by a
single shock but rather requires a mixture of both J-type and C-type
shocks (Burton et al. 1990, Wang \& Scoville 1992).

With such overwhelming evidence for an interaction, it may at first
seem puzzling why we have detected so few masers toward
IC\thinspace{443}. The 6 masers in Table 3 are all found within clump
``G'', one of a series of broad-line molecular clumps first imaged in
the CO line by DeNoyer (1979) and continued by Huang et al. (1986).
Clump G or IC\thinspace{443}G is located near the center of the
optical remnant but forms the western-most edge of a ring of shocked
molecular gas that can be traced in the lines of CO, HCO$^{+}$ and
H$_2$. The systematic variation of the peak velocity with position
prompted Dickman et al. (1992) and van Dishoeck et al. (1993) to model
the shocked gas as a tilted molecular ring, with the shock in
IC\thinspace{443}G transverse to the line of sight. Additional
evidence for this shock geometry comes from $^{12}$CO emission lines and
main line OH absorption lines, both of which show evidence for
pre-shock and post-shock gas (e.g. DeNoyer \& Frerking 1981).  In
IC\thinspace{443}B where the shock is thought to be mostly along the
line of sight to the observer, there is a clear separation in velocity
between the shocked and accelerated gas (the broad asymmetric
profiles) and the unshocked ambient gas (narrow profiles) (van
Dishoeck et al. 1993). In IC\thinspace{443}G the broad line
($\Delta$V$\simeq30$ km s$^{-1}$) remains but there is a narrow
self-absorption feature superimposed on emission lines like $^{12}$CO
and HCO$^{+}$ at a velocity between $-$4 and $-$5 km s$^{-1}$ (White
et al. 1987, van Dishoeck et al. 1993, Tauber et al. 1994). Such 
line profiles are expected from projection effects for a transverse
shock (see Turner et al. (1992) for a contrary picture). The
self-absorption is then due to cooler foreground gas, presumably the
unshocked ambient gas whose velocity is the systemic velocity of the
remnant (White et al. 1987).

The presence of a transverse shock in IC\thinspace{443}G may have
important implications for the lack of masers seen elsewhere in
IC\thinspace{443}. In the OH(1720 MHz) survey paper by Frail et
al. (1996) it was noted that in all instances the OH masers were
detected at or near the systemic velocity of the supernova remnant.
This was the case even when it was clear that the masers distributed
around a remnant arose from regions where the shock was propagating
into gas with very different properties (e.g. CTB 37A). We note that
the velocities of our masers in IC\thinspace{443}G are also clustered
close to the velocity of the the ambient gas ($\sim-$5 km s$^{-1}$).
In Frail et al. (1996) we suggested that this was due to the fact
that, similar to the case of stellar H$_2$O masers, the path of
maximum amplification occurs when the acceleration produced by the
shock is transverse to the line of sight. It is along this direction
that the velocity gradient is minimized and the largest coherent
pathlengths needed for bright maser emission are maintained. In
IC\thinspace{443}G we have for the first time clear evidence for the
existence of a transverse shock and thus compelling proof that
tangential amplification is important in the excitation of the OH(1720
MHz) masers. The lack of masers elsewhere in IC\thinspace{443}, where
the shock geometry is less favorable indirectly supports this
hypothesis. An alternate explanation for why the OH(1720 MHz) maser
velocities are close to the systemic velocity of the remnant is that
they are excited ahead of the shock in the unshocked ambient gas.
X-rays propagating from the shock into the dense molecular material
can form an X-ray dissociation region (XDR, Maloney, Hollenbach \&
Tielens 1996) with the densities and temperatures needed to invert the
1720 MHz line of OH. However, Burton et al. (1990) showed that the
luminosity of the infrared lines in IC\thinspace{443} such as [OI] are
too large to be explained by X-ray heating and that a shock model was
preferred.

IC\thinspace{443}G has been studied in a number of millimeter and
submillimeter lines by several groups (White et al. 1987, Ziurys et
al. 1989, Turner \& Lubowich 1991, Turner et al. 1992, van Dishoeck et
al. 1993, Tauber et al. 1994), allowing for accurate physical
parameters to be determined. At the highest angular resolution
provided by interferometric observations ($\sim$5\arcsec)
IC\thinspace{443}G breaks into two sub-clumps called GI and GII by van
Dishoeck et al. (1993) (G1 and G2 by Tauber et al. 1994). Five of the
six masers in Table 3 lie near the peak of GI, while our weakest maser
is closer to GII. The self-absorption of CO and other molecules in GII
was used by van Dishoeck et al. (1993) to constrain the gas kinetic
temperature and density of the ambient (unshocked) gas of T$_k$=10-20
K and $n_{\rm{H_2}}\sim$10$^4$ cm$^{-3}$. The parameters for the
postshock gas are more controversial and a range of densities and
temperatures seem to be present. Ziurys et al. (1989) used NH$_3$ to
derive T$_k$=33 K and argued that $n_{\rm{H_2}}>10^{5}$ cm$^{-3}$
because they detected HCO$^{+}$. A detection of H$_2$CO by Turner \&
Lubowich (1991) suggests much higher temperatures (T$_k>300$ K) and
densities ($n_{\rm{H_2}}\simeq10^7$ cm$^{-3}$). The multitransitional
studies by Turner et al. (1992) and van Dishoeck et al. (1993)
required at least two components to fit the observations of molecules
in IC\thinspace{443}G. The parameters of van Dishoeck et al. (1993)
fits give $n_{\rm{H_2}}\simeq10^5$ cm$^{-3}$, T$_k\sim80$ K for the
low density component and $n_{\rm{H_2}}\simeq3\times10^6$ cm$^{-3}$,
T$_k\sim{200}$ K for the high density component. Turner et al. (1992)
give similar values.

Based on these density and temperature measurements it is possible to
relate our observations to the Elitzur (1976) model which explains the
inversion of the OH(1720 MHz) maser line. With T$_k$=10-20 K it is
unlikely that the OH is excited in the ambient gas, since the pump
efficiency is much reduced below 25 K and fails to function below 15
K. It is equally improbable that the inversion is occurring in the high
density component of IC\thinspace{443}G. At T$_k>200$ K the excitation
of the 1612 MHz satellite line begins to dominate over 1720 MHz and at
$n_{\rm{H_2}}\gg{10}^5$ cm$^{-3}$ collisional de-excitation
depopulates the inversion (see also Pavlakis \& Kylafis 1996a). 
This leaves the low density component
($n_{\rm{H_2}}\simeq10^5$ cm$^{-3}$, T$_k\sim80$ K), which according
to Elitzur (1976) has the necessary density and temperature capable of
exciting the 1720 MHz masers. With more accurate modeling it should be
possible to predict the strength of the inversion and the size of the
masing regions in IC\thinspace{443}G, as was done approximately for
W\thinspace{28} (Frail et al. 1994).

\subsection{OH(1720 MHz) As A Tracer of Cosmic Ray Acceleration Sites}

Supernova remnant shocks have long been proposed as the acceleration
sites for cosmic rays (see reviews by Blandford \& Eichler 1987,
Biermann 1997) but direct observational proof is still
lacking. Arguments based on energetics and spectra demonstrate in a
self-consistent way that SNRs can provide the bulk of cosmic rays seen
above the ``knee'' near 5$\times{10}^{15}$ eV. For example, Duric et
al. (1995) have shown the cosmic rays in M\thinspace{33} can be
maintained by the observed SNR population. Moreover, the non-thermal
spectrum of the SNRs in our Galaxy (produced by synchrotron emission
from relativistic electrons) agree with the observed cosmic ray
spectrum after accounting for propagation effects. Powerful support of
this argument comes from the recent detection of X-ray synchrotron
emission from SN\thinspace{1006} (Koyama et al. 1995), as it
demonstrates that SNRs can accelerate electrons with the correct
spectrum to energies $\geq$200 Tev (Reynolds 1996 but see Mastichiadis
\& de Jager 1996 for a different viewpoint).


While it may never be possible to detect cosmic rays from SNRs
directly, collisions between cosmic rays and the atomic or molecular
gas in the ISM produce $\pi^\circ$ muons, which in turn decay to gamma
rays. Indeed, the distribution of the diffuse component of high energy
gamma rays ($>$100 MeV) follows the galactic distribution of gas
tracers like CO and HI nicely (Bloemen 1989). Embedded in the diffuse
emission are discrete gamma ray sources for which identification has
been more problematic. Early identifications with the COS-B satellite
included young pulsars such as Vela and the Crab and the quasar
3C\thinspace{273} (Swanenberg et al. 1981). The EGRET instrument on
the Compton Gamma Ray Observatory has continued to find more young
pulsars (Nel et al. 1996) and several active galactic nuclei, but a
stubborn number of the unresolved EGRET sources remain
unidentified. The second EGRET catalog and its supplement have
increased the original 22 COS-B sources with
$\vert{b}\vert\leq{10}^\circ$ to a total of 46 (Thompson et al. 1995,
1996).

The majority of these unidentified gamma ray sources are likely nearby
neutron stars, but a small subset could originate when cosmic rays,
accelerated in a supernova remnant shock, encounter the dense
molecular gas into which the remnant is expanding into and
compressing. This suggestion was first made for several of the COS-B
sources by Montmerle (1979), Morfill, Forman \& Bignami (1984) and
Pollock (1985). More recently both Sturner \& Dermer (1995) and
Esposito et al. (1996) have found examples of unidentified EGRET
sources which are at the same position as galactic SNRs with a low
probability of chance coincidence. In particular, Esposito et
al. (1996) have looked at a sample of 14 radio-bright SNRs and
found 5 candidates: W\thinspace{28}, W\thinspace{44},
IC\thinspace{443}, Monoceros and $\gamma$\thinspace{Cyg}. The EGRET
sources in Monoceros and $\gamma$\thinspace{Cyg} are both found
interior to the remnant and are likely young neutron stars.  Brazier
et al. (1996) have recently detected an X-ray source within the EGRET
error circle of $\gamma$\thinspace{Cyg} with characteristics similar
to Geminga. They argue that the gamma ray emission is from a young,
radio-quiet pulsar.

The three remaining SNRs also have known pulsars in their vicinity
(PSRs B0611+22, B1853+01 and B1758$-$23), two of which are within an
EGRET error circle. We cannot completely discount that the gamma ray
emission toward these SNRs originates from the pulsars (or a wind
nebula); however, searches have been made for pulsed high energy
emission without success (Nel et al. 1996). Furthermore, it is
noteworthy that the three remaining SNR/EGRET coincidences are also
our three best examples of a SNR/molecular cloud interaction, as traced by the
detection of the OH(1720 MHz) maser transition. Figure 10 shows the
positions of the 1720 MHz masers and the EGRET error circles overlaid
on the radio continuum.  Although the errors in
position are relatively large, the EGRET sources are located near the
edges of W\thinspace{28}, W\thinspace{44}, IC\thinspace{443}, not
inside them. Moreover, for W\thinspace{28} and W\thinspace{44} the
EGRET sources coincide with the quadrants where most of the masers are
located. For W\thinspace{28} this is the northeastern edge and for
W\thinspace{44} it is the eastern edge of the SNR. In the case of
IC\thinspace{443} the maser emission is less prominent (see \S{4.3}
for an explanation) but the EGRET source is found on the eastern edge
of the SNR where abundant neutral and molecular tracers indicate that
a strong shock along our line of sight is encountering a molecular
cloud (Braun \& Strom 1986, Burton et al. 1988, van Dishoeck, Jansen
\& Phillips 1993).

The correlation of the EGRET sources with SNRs, in and of itself, is
not a compelling proof that gamma rays originate as decay products
from the collision of cosmic rays with the gas surrounding an
SNR. However, the detection of the OH(1720 MHz) masers demonstrates
the existence of gas at a density between 10$^4$-10$^5$ cm$^{-3}$
which is in close proximity to the sites where the cosmic rays are
proposed to be accelerated. These density enhancements are expected to
produce local hot spots of gamma ray emission above the background
Drury, Aharonian \& V\"olk 1994). The detection of the OH(1720 MHz)
maser transition is much stronger evidence that a SNR is interacting
with a molecular cloud than the detection of HI or CO clouds towards
the SNR (e.g. Huang \& Thaddeus 1986). Without a clear shock signature
such detections have a high probability of being line of sight
coincidences. On the other hand, since the OH(1720 MHz) line is
thought to be collisionally excited by H$_2$ molecules heated by a
shock (Elitzur 1976), it is an unambiguous tracer of such interactions.


The OH(1720 MHz) line not only supports the cosmic ray model for the
origin of these EGRET sources as advocated by Sturner \& Dermer (1995)
and Esposito et al. (1996), it provides a useful target list for
future observations with more sensitive GeV satellites or instruments
at TeV energies where the background is much reduced (Kifune
1995). There are a total of 17 SNRs towards which OH(1720 MHz) masers
have been detected (Green et al. 1997 and references therein).  
The flux density of the gamma rays produced via
$\pi^\circ$ decay depends directly on the fraction of the supernova
blast energy E$_\circ$ that goes into accelerating the cosmic rays and
the density of the gas, and inversely on the square of the distance to
the SNR (Aharonian, Drury \& V\"olk 1994, Drury et al. 1994).
W\thinspace{28}, W\thinspace{44}, IC\thinspace{443} are the closest
SNRs. The remaining 14 SNRs all are more distant than 5 kpc and thus
we would not expect these SNR detected as discrete EGRET sources given
its sensitivity (Drury et al. 1994). However, they are important
candidates for future searches and their detection at Gev or TeV
energies would be a convincing proof of cosmic ray acceleration
theories.

\section{Conclusions}

The observations presented here leave little doubt that the OH(1720
MHz) features that we have detected toward supernova remnants are
masers. Estimates of the temperature and densities based on thermal
molecular data agree with those used in the shock
excitation model of Elitzur (1976). The limits on the maser optical
depth and brightness temperatures are of the correct magnitude. The
distribution of the masers behind the leading edge of the shock, as
traced by the relativistic electrons, suggest that the lines originate
in postshock gas. Clear evidence for this also exists in the specific
case of \ic443, where the masers are positionally coincident with
shocked gas (i.e. IC\thinspace{443}G). We have also shown that the
maser velocities cluster around the systemic velocity of the remnant
and the parent molecular cloud into which the shock is propagating.
This, together with the existence of a transverse shock for \ic443,
has been used to argue that the tangential amplification is important
for the excitation of OH(1720 MHz) masers.

Despite this progress some uncertainty remains. The peculiar
kinematics of W\thinspace{28} do not fit this simple picture.  The
smooth gradient in velocity as one moves outward to the edge of the
W\thinspace{28} SNR, and the large increase in the velocity dispersion
at the edge, are difficult to understand. A molecular cloud
accelerated by the SNR toward the line-of-sight was offered as an
explanation but it is not a unique explanation nor is it an entirely
satisfactory one. A first-ever measurement of the strength of the
magnetic field in post-shock gas behind a supernova remnant has
allowed us to demonstrate the dominance of magnetic pressure in these
shocks. However, the details of the shock physics are still not
clear. Is the shock dissociative or do the H$_2$ and OH survive the
passage of the shock?  Likewise the physical parameters of the pre-
and post-shock gas are not well constrained except perhaps for
\ic443. Better theoretical modeling of the shock and maser excitation
process are needed. While it may be possible to infer the density and
temperature of the gas from the masers lines alone, observations of
millimeter and sub-millimeter lines should be used as an important
check on the OH(1720 MHz) maser models. Higher angular observations
with full Stokes parameters should also be pursued to measure the true
intrinsic size of the OH(1720 MHz) masers (and hence T$_b$ and $\tau$)
and to check that the derivation of magnetic fields from the Stokes
{\bf V} profile fitting is not confused by unrelated, but overlapping
features in the arcsecond beam.

\acknowledgements

This research has made use of NASA's Astrophysics Data System Abstract
Service (ADS) and the Simbad database, operated at CDS, Strasbourg,
France.

\clearpage
\begin{figure}

\caption{A 327 MHz radio continuum image (Frail, Kulkarni \& Vasisht 1993) 
of the W\thinspace{28} SNR with the
location of the various regions of OH(1720 MHz) emission
concentrations indicated.  The beam size is 55\arcs by 41\arcs at
position angle 69\arcdeg, and the grey scale at the top is in mJy beam$^{-1}$.
These regions are shown in more detail in Fig. 4.}

\caption{A 1442 MHz radio continuum image (Giacani \etal 1997) 
of the W\thinspace{44} SNR with the
location of the various regions of OH(1720 MHz) emission
concentrations indicated. The beam size is 15\arcs, and the grey scale 
at the top is in mJy beam$^{-1}$.
These regions are shown in more detail in Fig. 5.}

\caption{A 330 MHz radio continuum image (Kassim, private communication)
of the IC\thinspace{443} SNR with the location of the OH(1720 MHz) 
emission as indicated. The beam size is 80\arcs by 70\arcs at position
angle 50\arcdeg, and the grey scale at the top is in mJy beam$^{-1}$.
This region is shown in more detail in Fig. 6.}

\caption{Individual regions of OH(1720 MHz) emission concentrations
(a-f) in the W\thinspace{28} SNR. For each region the radio continuum
contours are plotted and labeled in mJy beam$^{-1}$. Individual OH(1720
MHz) masers are also plotted with exaggerated sizes to reflect the
apparent OH(1720 MHz) spot sizes as determined by our fitting process
(see \S{2}). The size of the ellipse is ten times the angular
diameters given in Table 2. Unresolved features are indicated by a
filled circle.}

\caption{Identical to Fig. 4. except for the W\thinspace{44} SNR.}

\caption{Identical to Fig. 4. except for the IC\thinspace{443} SNR.}

\caption{The distribution of maser velocities (LSR) across
W\thinspace{28}. The geometric center of W\thinspace{28} is taken to
be at $\alpha=17^h~57^m~47^s$,
$\delta=-23^\circ~{20}^\prime~{24}^{\prime\prime}$ (B1950). Individual
maser features are plotted with open circles, the diameter of which is
proportional to the logarithm of their peak flux density.}

\caption{Stokes parameter spectra for maser feature 3 (see Table 2) in
the W\thinspace{28} SNR.  Shown with the {\bf V} spectrum, as a dashed line,
is the frequency derivative of the {\bf I} spectrum, scaled by the 
magnetic field strength. }

\caption{Identical to Fig. 8, except for maser feature 11 (see Table 3) in the
W\thinspace{44} SNR.}

\caption{The location of discrete gamma-ray sources with respect to 
W\thinspace{28}, W\thinspace{44} and IC\thinspace{443} as detected by
the EGRET experiment on the Compton Gamma Ray Observatory. Large
ellipses are the EGRET error circles, smaller circles are the regions
of OH(1720 MHz) emission.}

\end{figure}
\clearpage 

\end{document}